\documentclass[aps,preprintnumbers,superscriptaddress,showpacs]{revtex4}
\usepackage{epsfig}
\usepackage{psfrag}
\usepackage{amsfonts}
\usepackage{graphicx}
\usepackage{dcolumn}
\usepackage{bm}

\begin{document}

\title{Holographic Schwinger effect with a moving D3-brane}

\author{Zi-qiang Zhang}
\email{zhangzq@cug.edu.cn} \affiliation{School of mathematics and
physics, China University of Geosciences(Wuhan), Wuhan 430074,
China}

\author{De-fu Hou}
\email{houdf@mail.ccnu.edu.cn} \affiliation{Key Laboratory of
Quark and Lepton Physics (MOE), Central China Normal University,
Wuhan 430079,China}

\author{Gang Chen}
\email{chengang1@cug.edu.cn} \affiliation{School of mathematics
and physics, China University of Geosciences(Wuhan), Wuhan 430074,
China}

%%%%%%%%%%%%%%%%%%%%%%%%%%%%%%%%%%%%%%%%
\begin{abstract}
We study the Schwinger effect with a moving D3-brane in a
$\mathcal{N}$=4 SYM plasma with the aid of AdS/CFT correspondence.
We discuss the test particle pair moving transverse and parallel
to the plasma wind respectively. It is found that for both cases
the presence of velocity tends to increase the Schwinger effect.
In addition, the velocity has a stronger influence on the
Schwinger effect when the pair moves transverse to the plasma wind
rather than parallel.

\end{abstract}
\pacs{11.25.Tq, 11.15.Tk, 11.25-w}

\maketitle
%%%%%%%%%%%%%%%%%%%%%%%%%%%%%%%%%%%%%%%%
\section{Introduction}
Schwinger effect is an interesting phenomenon in quantum
electrodynamics (QED) \cite{JS}. The virtual particles can be
materialized and turn into real particles due to the presence of
an external strong electric field. The production rate $\Gamma$
has been studied for the case of weak-coupling and weak-field long
time ago \cite{JS}
\begin{equation}
\Gamma\sim exp({\frac{-\pi m^2}{eE}}),
\end{equation}
where $e$, $m$ and $E$ are the elementary electric charge, the
electron mass and the external electric field, respectively. In
this case, there is no critical field. Thirty years later, the
calculation of $\Gamma$ has been generalized to the case of
arbitary-coupling and weak-field regime \cite{IK}
\begin{equation}
\Gamma\sim exp({\frac{-\pi m^2}{eE}+\frac{e^2}{4}}).
\end{equation}
In this case, the exponential suppression vanishes when one takes
$eE_c=(4\pi/e^2)m^2\simeq137m^2$. However, the critical value
$E_c$ does not agree with the weak-field condition $eE\ll m^2$.
Thus, it seems to be an obstacle to evaluate the critical field
under the weak-field condition. One step further, we don't know
whether the catastrophic decay really occur or not.

Interestingly, there exists a critical value of the electric field
in string theory, and the results in string theory suggest that
the catastrophic vacuum decay can really occur in some cases
\cite{ES,CB}. It is well known that the string theory can dual to
the gauge theory through the AdS/CFT correspondence
\cite{Maldacena:1997re,Gubser:1998bc,MadalcenaReview}. Therefore,
it is of great interest to study the Schwinger effect in a
holographic setup. On the other hand, the Schwinger effect might
not be intrinsic only for QED but rather a general feature for
QFTs coupled to an U(1) gauge field. Recently, Semenoff and
Zarembo argued \cite{GW} that one can realize a $\mathcal N=4$ SYM
theory system that coupled with an U(1) gauge field through the
Higgs mechanism. In this approach, the production rate of the
fundamental particles, at large $N$ and large 't Hooft coupling
$\lambda\equiv Ng_{YM}^2$, has been evaluated as
\begin{equation}
\Gamma\sim
exp[-\frac{\sqrt{\lambda}}{2}(\sqrt{\frac{E_c}{E}}-\sqrt{\frac{E}{E_c}})^2],\qquad
E_c=\frac{2\pi m^2}{\sqrt{\lambda}},
\end{equation}
intriguingly, the critical electric field $E_c$ completely agrees
with the Dirac-Born-Infeld (DBI) result. Motivated by \cite{GW},
there are many attempts to address the Schwinger effect in this
direction. For instance, the potential analysis for the pair
creation is studied in \cite{YS2}. The universal aspects of the
Schwinger effect in general backgrounds with an external
homogeneous electric field are analyzed in \cite{YS}. The
Schwinger effect in confining backgrounds is discussed in
\cite{YS1}. The Schwinger effect with constant electric and
magnetic fields has been investigated in \cite{SB,YS3}. The
consequences of the Schwinger effect for conductivity has been
addressed in \cite{SC}. Other related results can be found, for
example in \cite{JA,KH1,DD,WF,MG,XW,ZQ}. For reviews on this
topic, see \cite{DK1} and references therein.

Now we would like to study the influence of velocity on the
Schwinger effect. The motivation comes from the experiments: the
particles are not produced at rest but observed moving with
relativistic velocities through the medium, so the effect of
velocity should be taken into account. For that reason, the
velocity effect on some quantities has been studied. For example,
the influence of velocity on the Im$V_{Q\bar{Q}}$ is investigated
in \cite{MAL}.  The velocity effect on the entropic force is
analyzed in \cite{KBF}. As the Schwinger effect with a static
D3-brane has been discussed in \cite{YS2}, it is also of interest
to extend this study to the case of a moving D3-brane. In this
paper, we would like to see how velocity affects the Schwinger
effect. This is the purpose of the present work.

The paper is organized as follows. In the next section, we briefly
review the Schwarzschild $AdS_5\times S^5$ background and boost
the frame in one direction. In section 3, we perform the potential
analysis for the pair moving transverse and parallel to the plasma
wind respectively.  Also, we calculate the critical electric field
by DBI action. The last part is devoted to conclusion and
discussion.

\section{setup}

Let us briefly introduce the Schwarzschild $AdS_5\times S^5$
background. The metric of this black hole in the Lorentzian
signature is given by \cite{GT}
\begin{eqnarray}
ds^2
=\frac{r^2}{R^2}[-f(r)dt^2+d\vec{x}^2]+\frac{R^2}{r^2}f(r)^{-1}dr^2+R^2d\Omega_5^2,\label{metric}
\end{eqnarray}
with
\begin{equation}
f(r)=1-(\frac{r_h}{r})^4,
\end{equation}
where $R$ is the AdS space radius, $\vec{x}$ denotes the spatial
directions of the space time, r stands for the radial coordinate
of the geometry. The horizon is located at $r=r_h$ and the
temperature is
\begin{equation}
T=\frac{r_h}{\pi R^2}.
\end{equation}

To make the D3-brane (or particle pair) moving, we assume that the
plasma is at rest and the frame is moving with a velocity in one
direction, i.e., we boost the frame in the $x_3$ direction so that
\cite{KBF}
\begin{equation}
dt=dt^\prime cosh\beta-dx_3^\prime sinh\beta, \qquad
dx_3=-dt^\prime sinh\beta+dx_3^\prime cosh\beta,\label{tr}
\end{equation}
where $\beta$ is the velocity (or rapidity).

Substituting (\ref{tr}) into (\ref{metric}) and removing the
primes, we have the boosted metric as
\begin{eqnarray}
ds^2&=&-\frac{r^2}{R^2}[f(r)cosh^2\beta-sinh^2\beta]dt^2+\frac{r^2}{R^2}(dx_1^2+dx_2^2)+\frac{r^2}{R^2}[cosh^2\beta-f(r)sinh^2\beta]dx_3^2\nonumber\\&-&2sinh\beta
cosh\beta\frac{r_h^4}{r^2R^2}dtdx_3+\frac{R^2}{r^2}f(r)^{-1}dr^2+R^2d\Omega_5^2.\label{metric1}
\end{eqnarray}

Note that for $\beta=0$ in (\ref{metric1}) the metric of
(\ref{metric}) is reproduced.

\section{potential analysis}

In this section, we follow the calculations of \cite{YS2} to study
the Schwinger effect with the metric (\ref{metric1}). Generally,
to analyze the moving case, one should consider different
alignments for the particle pair with respect to the plasma wind,
including transverse ($\theta=\pi/2$), parallel ($\theta=0$), or
arbitrary direction ($\theta$). In the present work, we discuss
two cases: $\theta=\pi/2$ and $\theta=0$.

\subsection{Transverse to the wind $(\theta=\pi/2)$}

We now analyze the system perpendicularly to the wind, the
coordinate is parameterized by
\begin{equation}
t=\tau, \qquad x_1=\sigma,\qquad x_2=0,\qquad x_3=0,\qquad
r=r(\sigma),\label{par1}
\end{equation}
where the test particles (quark and anti-quark) are located at
$x_1=-\frac{x}{2}$ and $x_1=\frac{x}{2}$ with $x$ the
inter-distance.

The Nambu-Goto action is
\begin{equation}
S=T_F\int d\tau d\sigma\mathcal L=T_F\int d\tau d\sigma\sqrt{g},
\label{S}
\end{equation}
where $T_F=\frac{1}{2\pi\alpha^\prime}$ is the string tension and
$\alpha^\prime$ is related to $\lambda$ by
$\frac{R^2}{\alpha^\prime}=\sqrt{\lambda}$. Here $g$ is the
determinant of the induced metric with
\begin{equation}
g_{\alpha\beta}=g_{\mu\nu}\frac{\partial
X^\mu}{\partial\sigma^\alpha} \frac{\partial
X^\nu}{\partial\sigma^\beta},
\end{equation}
where $g_{\mu\nu}$ is the metric, $X^\mu$ stands for the target
space coordinates. The world sheet is parameterized by
$\sigma^\alpha$ with $\alpha=0,1$.

From (\ref{metric1}) and (\ref{par1}) one gets the induced metric
as follows
\begin{equation} g_{00}=\frac{r^2}{R^2}f(r)cosh^2\beta-\frac{r^2}{R^2}sinh^2\beta, \qquad
g_{11}=\frac{r^2}{R^2}+\frac{R^2}{f(r)r^2}\dot{r}^2.
\end{equation}
with $\dot{r}=dr/d\sigma$.

This leads to a lagrangian density
\begin{equation}
\mathcal L=\sqrt{a(r)+b(r)\dot{r}^2},\label{L}
\end{equation}
with
\begin{eqnarray}
&a(r)&=\frac{r^4}{R^4}f(r)cosh^2\beta-\frac{r^4}{R^4}sinh^2\beta,\nonumber\\&b(r)&=cosh^2\beta-\frac{1}{f(r)}sinh^2\beta.
\end{eqnarray}

Now that $\mathcal L$ does not depend on $\sigma$ explicitly, so
the corresponding Hamiltonian is a constant
\begin{equation}
H=\mathcal L-\frac{\partial\mathcal
L}{\partial\dot{r}}\dot{r}=Constant.
\end{equation}

The boundary condition at $\sigma=0$ is
\begin{equation}
\dot{r}=0,\qquad  r=r_c\qquad (r_h<r_c<r_0)\label{con},
\end{equation}
which yields
\begin{equation}
\dot{r}=\frac{dr}{d\sigma}=\sqrt{\frac{a^2(r)-a(r)a(r_c)}{a(r_c)b(r)}}\label{dotr}.
\end{equation}
with
\begin{equation}
a(r_c)=\frac{r_c^4}{R^4}f(r_c)cosh^2\beta-\frac{r_c^4}{R^4}sinh^2\beta,\qquad
f(r_c)=1-(\frac{r_h}{r_c})^4.
\end{equation}

By integrating (\ref{dotr}), we can get the separate length of the
test particles on the probe brane as
\begin{equation}
x=\frac{2R^2}{r_0a}\int_1^{1/a}dy\sqrt{\frac{a(y_c)b(y)}{a^2(y)-a(y)a(y_c)}}\label{xx},
\end{equation}
with
\begin{equation}
a(y)=y^4f(y)cosh^2\beta-y^4sinh^2\beta,\qquad
a(y_c)=f(y_c)cosh^2\beta-sinh^2\beta,\qquad
b(y)=cosh^2\beta-\frac{1}{f(y)}sinh^2\beta,
\end{equation}
and
\begin{equation}
f(y)=1-\frac{b^4}{(ay)^4},\qquad f(y_c)=1-\frac{b^4}{a^4},
\end{equation}
where we have introduced the following dimensionless parameters
\begin{equation}
y\equiv\frac{r}{r_c},\qquad a\equiv\frac{r_c}{r_0},\qquad
b\equiv\frac{r_h}{r_0}.
\end{equation}

On the other hand, from (\ref{S}), (\ref{L}) and (\ref{dotr}), the
sum of Coulomb potential and energy can be written as
\begin{equation}
V_{CP+E}=2T_Fr_0a\int_1^{1/a}dy\sqrt{\frac{a(y)b(y)}{a(y)-a(y_c)}}.\label{en}
\end{equation}

Next, we calculate the critical field. The DBI action is given by
\begin{equation}
S_{DBI}=-T_{D3}\int
d^4x\sqrt{-det(G_{\mu\nu}+\mathcal{F}_{\mu\nu})}\label{dbi},
\end{equation}
where $T_{D3}$ is the D3-brane tension
\begin{equation}
T_{D3}=\frac{1}{g_s(2\pi)^3\alpha^{\prime^2}}.
\end{equation}

By virtue of (\ref{metric1}), the induced metric $G_{\mu\nu}$
becomes
\begin{equation}
G_{00}=-\frac{r^2}{R^2}f(r)cosh^2\beta+\frac{r^2}{R^2}sinh^2\beta,
\qquad G_{11}= G_{22}=\frac{r^2}{R^2},\qquad
G_{33}=-\frac{r^2}{R^2}f(r)sinh^2\beta+\frac{r^2}{R^2}cosh^2\beta.
\end{equation}

Considering $\mathcal{F}_{\mu\nu}=2\pi\alpha^\prime F_{\mu\nu}$
\cite{BZ} and supposing that the electric field is turned on along
the $x^1$-direction \cite{YS2}, we have
\begin{equation}
G_{\mu\nu}+\mathcal{F}_{\mu\nu}=\left(
\begin{array}{cccc}
-\frac{r^2}{R^2}f(r)cosh^2\beta+\frac{r^2}{R^2}sinh^2\beta & 2\pi\alpha^\prime E & 0 & 0\\
 -2\pi\alpha^\prime E & \frac{r^2}{R^2} & 0 & 0 \\
 0 & 0 & \frac{r^2}{R^2} & 0\\
0 & 0 & 0 &
-\frac{r^2}{R^2}f(r)sinh^2\beta+\frac{r^2}{R^2}cosh^2\beta
\end{array}
\right),
\end{equation}
which yields
\begin{equation}
det(G_{\mu\nu}+\mathcal{F}_{\mu\nu})=\frac{r^4}{R^4}[cosh^2\beta-f(r)sinh^2\beta][(2\pi\alpha^\prime)^2E^2-\frac{r^4}{R^4}(f(r)cosh^2\beta-sinh^2\beta)].\label{det}
\end{equation}

Substituting (\ref{det}) into (\ref{dbi}) and making the D3-brane
located at $r=r_0$, one gets
\begin{equation}
S_{DBI}=-T_{D3}\frac{r_0^2}{R^2}\int d^4x
\sqrt{[cosh^2\beta-f(r_0)sinh^2\beta][\frac{r_0^4}{R^4}(f(r_0)cosh^2\beta-sinh^2\beta)-(2\pi\alpha^\prime)^2E^2]}\label{dbi1},
\end{equation}
with
\begin{equation}
f(r_0)=1-(\frac{r_h}{r_0})^4=1-b^4.
\end{equation}

Obviously,
\begin{equation}
cosh^2\beta-f(r_0)sinh^2\beta\geq0.
\end{equation}

So to avoid (\ref{dbi1}) being ill-defined one needs only
\begin{equation}
\frac{r_0^4}{R^4}(f(r_0)cosh^2\beta-sinh^2\beta)-(2\pi\alpha^\prime)^2E^2\geq0,\label{ec}
\end{equation}
which leads to
\begin{equation}
E\leq T_F\frac{r_0^2}{R^2}\sqrt{f(r_0)cosh^2\beta-sinh^2\beta}.
\end{equation}

Thus, the critical field $E_c$ is obtained
\begin{equation}
E_c=T_F\frac{r_0^2}{R^2}\sqrt{f(r_0)cosh^2\beta-sinh^2\beta},\label{ec1}
\end{equation}
one can see that $E_c$ depends on the velocity as well as the
temperature.

Now we are ready to calculate the total potential. As a matter of
convenience, we introduce a dimensionless quantity here
\begin{equation}
\alpha\equiv\frac{E}{E_c}. \label{afa}
\end{equation}

Then, from (\ref{xx}),(\ref{en}) and (\ref{afa}) one finds the
total potential
\begin{eqnarray}
V_{tot}(x)&=&V_{CP+E}-Ex\nonumber\\&=&2T_Fr_0a\int_1^{1/a}dy\sqrt{\frac{a(y)b(y)}{a(y)-a(y_c)}}\nonumber\\&-&
\frac{2T_F\alpha
r_0}{a}\sqrt{f(r_0)cosh^2\beta-sinh^2\beta}\int_1^{1/a}dy\sqrt{\frac{a(y_c)b(y)}{a^2(y)-a(y)a(y_c)}}
\label{V}.
\end{eqnarray}

We have checked that the $V_{tot}(x)$ in the Schwarzschild
$AdS_5\times S^5$ background with a static D3-brane can be derived
from (\ref{V}) if we neglect the effect of velocity by plugging
$\beta=0$ in (\ref{V}), as expected.

According to the potential analysis in \cite{YS2}, there exist a
critical value of the electric field
$E_c=\frac{T_Fr_0^2}{R^2}\sqrt{1-\frac{r_h^4}{r_0^4}}$.  When
$E<E_c$, the potential barrier exists and the pair creation can be
described as a tunneling phenomenon. When $E>E_c$, the potential
barrier vanishes and the vacuum becomes unstable.

Let us discuss results. To compare with the case in \cite{YS2}, we
set $\frac{R^2}{r_0}=T_Fr_0=1$ and $b=0.5$ here. In Fig.1 we plot
the total potential $V_{tot}(x)$ as a function of the distance $x$
with two fixed velocity for different $\alpha$. The left is
plotted for $\beta=0.1$ and the right is for $\beta=0.8$. From the
figures, we can indeed see that there exists a critical electric
field at $\alpha=1$($E=E_c$) for different $\beta$, consistently
with the findings of \cite{YS2}.

\begin{figure}
\centering
\includegraphics[width=0.4\textwidth,bb=0 0 800 600]{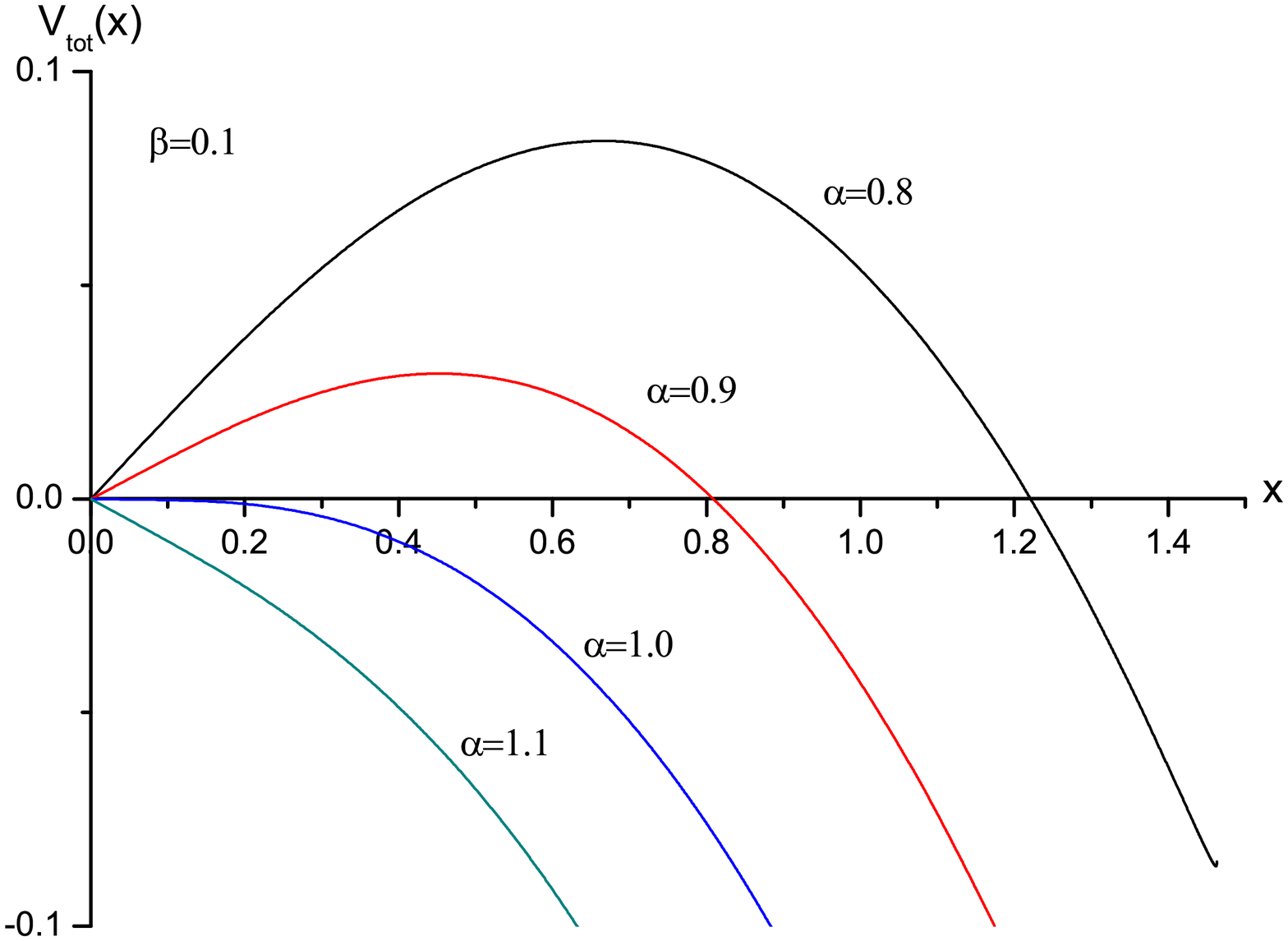}
\includegraphics[width=0.4\textwidth,bb=0 0 800 600]{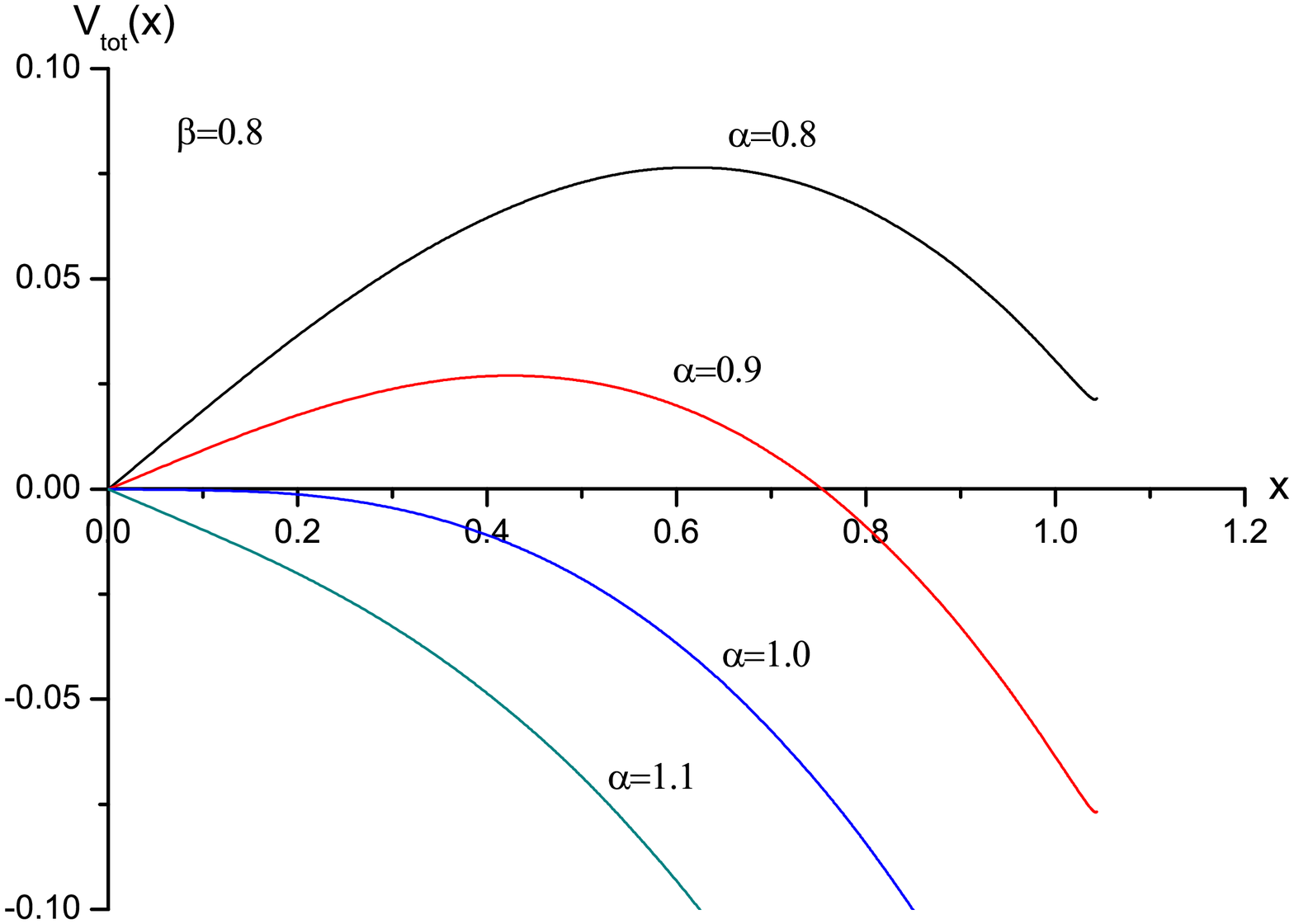}
\caption{$V_{tot}(x)$ versus $x$. Left: $\beta=0.1$; Right:
$\beta=0.8$. In all of the plots from top to bottom
$\alpha=0.8,0.9,1.0,1.1$, respectively.}
\end{figure}

To show the effect of velocity on the potential barrier we plot
$V_{tot}(x)$ against $x$ with $\alpha=0.9$ for different $\beta$
in the left panel of Fig.2. We can see that by increasing $\beta$
the height and width of the potential barrier both decrease. As we
know the higher the potential barrier, the harder the produced
pair escape to infinity. Therefore, the presence of velocity tends
to increase the Schwinger effect.

To study how the velocity affects the value of $E_c$ we plot
$V_{tot}(x)$ versus $x$ for different $\beta$ at $\alpha=1$ in the
right panel of Fig.2. From the figures, we can see that the
barrier vanishes for each plot implying the vacuum becomes
unstable. Also, by increasing $\beta$ the height of the plot
decreases, in agreement with the analysis of (\ref{ec1}).
Actually, that the barrier of each plot disappears at $\alpha=1$
can be strictly proved, i.e, one can calculate the
$\frac{dV_{tot}(x)}{dx}$ at $x$ as
\begin{equation}
\frac{dV_{tot}}{dx}|_{x=0}=(1-\alpha)T_F\sqrt{f(r_0)cosh^2\beta-sinh^2\beta}.
\end{equation}

\begin{figure}
\centering
\includegraphics[width=0.4\textwidth,bb=0 0 800 600]{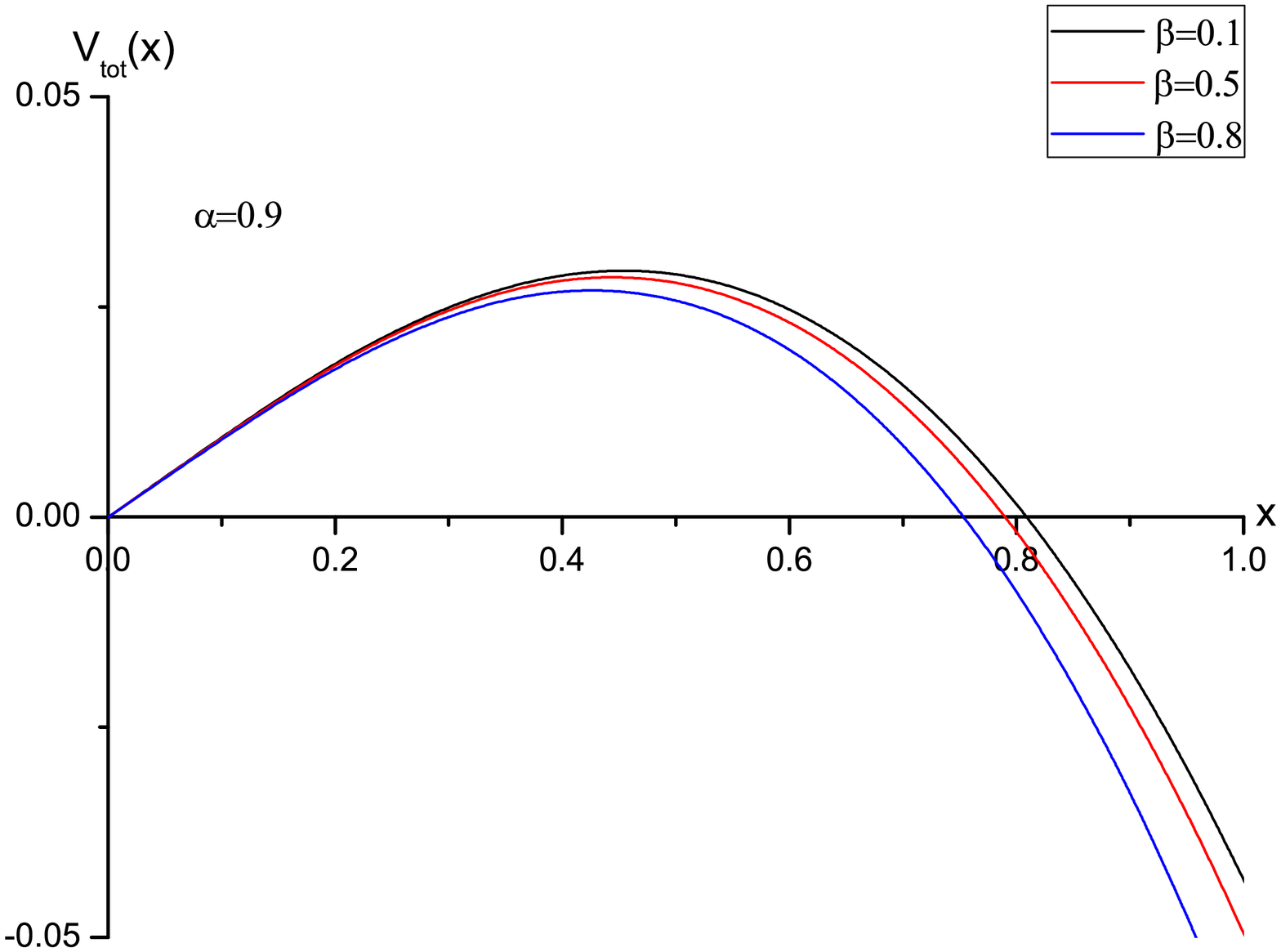}
\includegraphics[width=0.4\textwidth,bb=0 0 800 600]{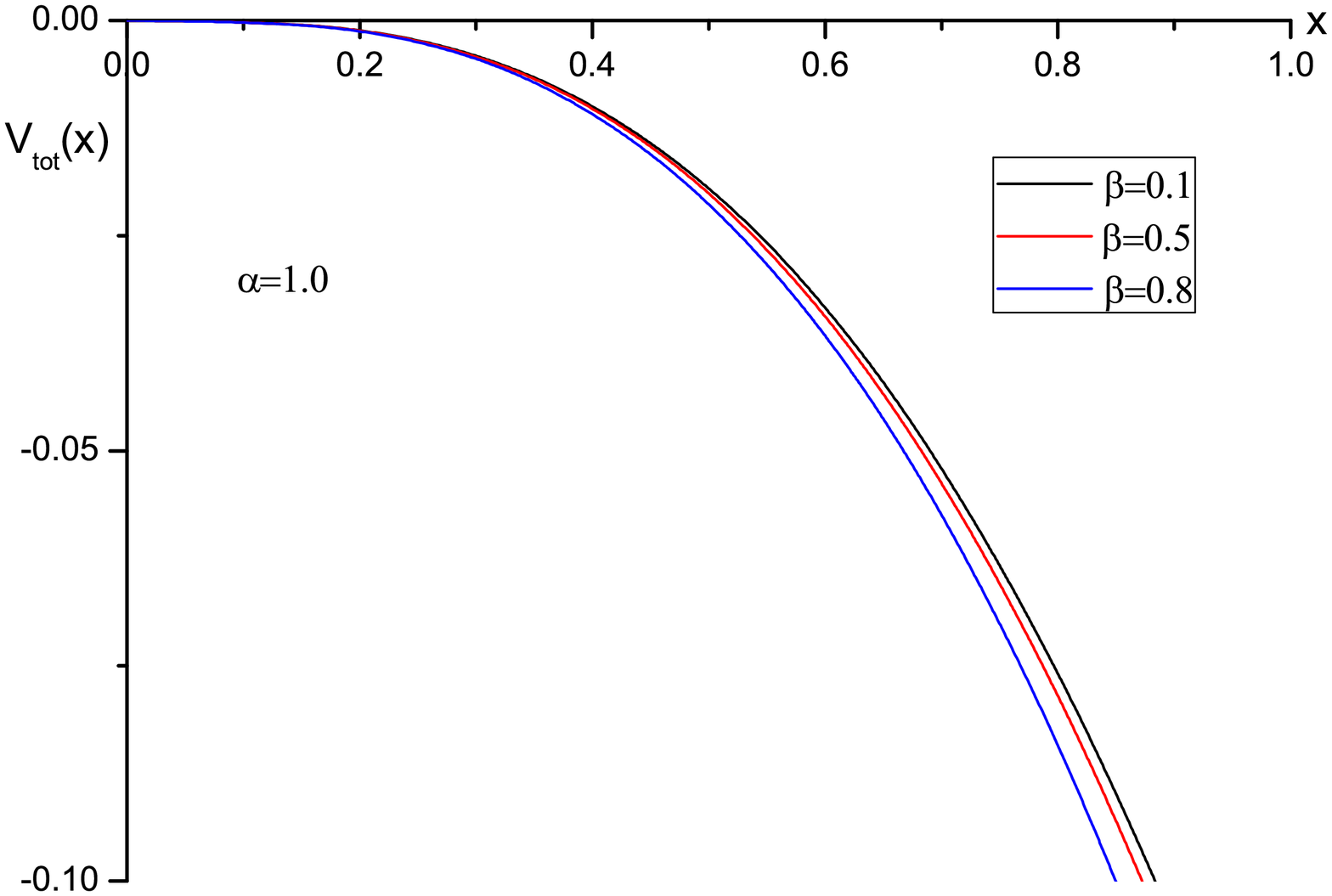}
\caption{$V_{tot}(x)$ versus $x$. Left: $\alpha=0.9$; Right:
$\alpha=1$. In all of the plots from top to bottom
$\beta=0.1,0.5,0.8$, respectively.}
\end{figure}

\subsection{Parallel to the wind $(\theta=0)$}

We now discuss the system parallel to the wind. The coordinate is
parameterized by
\begin{equation}
t=\tau, \qquad x_1=0,\qquad x_2=0,\qquad x_3=\sigma,\qquad
r=r(\sigma).\label{par}
\end{equation}
where the test particles are located at $x_3=-\frac{x}{2}$ and
$x_3=\frac{x}{2}$, respectively.

The next analysis is almost similar to the previous subsection. So
we here show the final results. The total potential is given by
\begin{eqnarray}
V_{tot}(x)&=&2T_Fr_0a\int_1^{1/a}dy\sqrt{\frac{A(y)B(y)}{A(y)-A(y_c)}}\nonumber\\&-&
\frac{2T_F\alpha
r_0}{a}\sqrt{(f(r_0)cosh^2\beta-sinh^2\beta)(cosh^2\beta-f(r_0)sinh^2\beta)}\int_1^{1/a}dy\sqrt{\frac{A(y_c)B(y)}{A^2(y)-A(y)A(y_c)}}
\label{V1}.
\end{eqnarray}
with
\begin{equation}
A(y)=y^4[f(y)(sinh^4\beta+cosh^4\beta)-(1+f^2(y))sinh^2\beta
cosh^2\beta],\qquad
B(y)=cosh^2\beta-\frac{1}{f(y)}sinh^2\beta,
\end{equation}
and
\begin{equation}
A(y_c)=f(y_c)(sinh^4\beta+cosh^4\beta)-(1+f^2(y_c))sinh^2\beta
cosh^2\beta.
\end{equation}

\begin{figure}
\centering
\includegraphics[width=0.4\textwidth,bb=0 0 800 600]{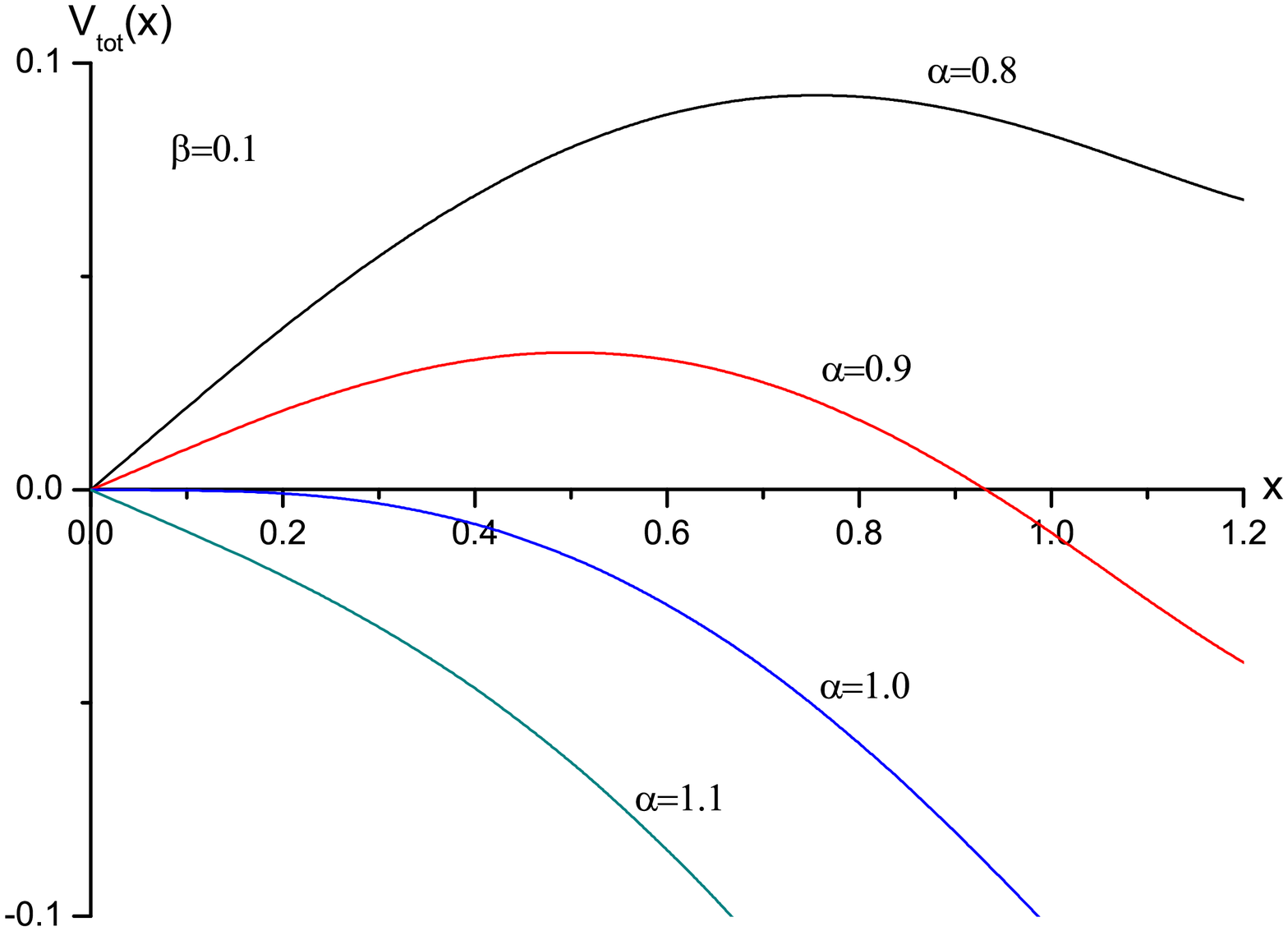}
\includegraphics[width=0.4\textwidth,bb=0 0 800 600]{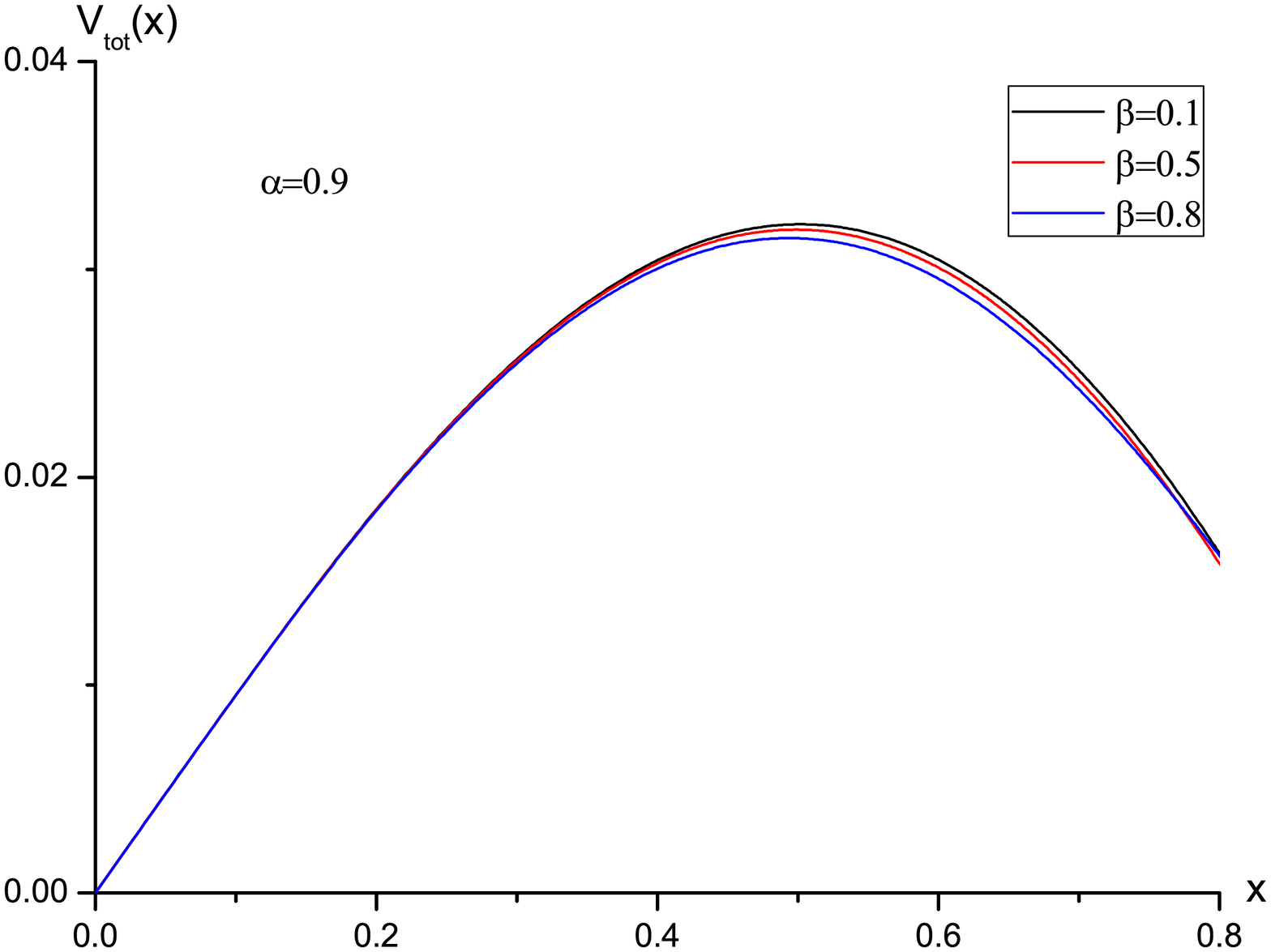}
\caption{$V_{tot}(x)$ versus $x$. Left: $\beta=0.1$; from top to
bottom $\alpha=0.8,0.9,1.0,1.1$, respectively. Right:
$\alpha=0.9$, from top to bottom $\beta=0.1,0.5,0.8$,
respectively.}
\end{figure}

In Fig.3, we also plot $V_{tot}(x)$ as a function of $x$ for
$\theta=0$ in two cases. We can see that the behavior is similar
to the case of $\theta=\pi/2$. The only difference is that $\beta$
has a smaller influence on the Schwinger effect when the pair is
moving parallel to the plasma wind. Interestingly, the velocity
also has a smaller influence on the imaginary potential \cite{MAL}
and the entropic force \cite{KBF} when the pair moves parallel to
the wind rather than transverse.

\section{conclusion and discussion}
In heavy ion collisions at LHC and RHIC, the produced pair is
moving through the medium with relativistic velocities. An
understanding of how some quantities are affected by the
velocities may be essential for theoretical predictions. In this
paper, we have investigated the influence of velocity on the
Schwinger effect at finite temperature from the AdS/CFT
correspondence. We have studied the pair moving transverse and
parallel to the plasma wind respectively. The potential analysis
for these backgrounds was presented. The value of the critical
electric field was obtained. It is shown that for both cases the
presence of velocity tends to increase the production rate. In
addition, the velocity has a stronger influence on the Schwinger
effect when the pair moves transverse to the plasma wind rather
than parallel.

Finally, it is interesting to mention that the holographic
Schwinger effect with a rotating D3-brane has been studied in
\cite{HXU} recently.

\section{Acknowledgments}

This work is partly supported by the Ministry of Science and
Technology of China (MSTC) under the ¡°973¡± Project no.
2015CB856904(4). Zi-qiang Zhang and Gang Chen are supported by the
NSFC under Grant no. 11475149. De-fu Hou is supported by the NSFC
under Grant no. 11375070 and 11521064.

%%%%%%%%%%%%%%%%%%%%%%%%%%%%%%%%%%%%%%%%

\end{document}